\documentclass[prl,twocolumn,aps]{revtex4}
\usepackage[dvips]{graphicx}
\usepackage{epsfig}
\usepackage{amsmath,amsfonts,amssymb,bm}
\usepackage{bm}
\usepackage[caption=false]{subfig}
\begin{document}
\title{Many-Body Delocalization in Strongly Disordered System with Long-Range Interactions: Finite Size Scaling}
\author{Alexander L. Burin}
\affiliation{Department of Chemistry, Tulane University, New
Orleans, LA 70118, USA}
\date{\today}
\begin{abstract}
The localization in a disordered system of $N$ interacting spins coupled by the long-range anisotropic interaction  $1/R^{\alpha}$ is investigated  using a finite size scaling in a $d=1$ -dimensional system for  $N=8, 10, 12, 14$.  The results supports the absence of localization in the infinite system at $\alpha<2d$ and a scaling of a critical energy disordering $W_{c} \propto N^{2d-\alpha}$ in agreement with the analytical theory suggesting the energy delocalization in the subset of interacting resonant pairs of spins as a precursor of the many-body delocalization.The spin relaxation rate $k$ dependence on disordering $k \propto W^{-2}$ has been revealed in the practically interesting case $\alpha=d$. This relaxation mechanism can be responsible for the  anomalous relaxation of quantum two level systems in amorphous solids. 
\end{abstract}

\maketitle


Localization - delocalization transition separates non-ergodic, reversible behavior from a chaotic, ergodic, irreversible regime and therefore it is in the focus of the scientific community since the concept of localization has been suggested for a single particle in a random external field \cite{AndersonClassic}. At present the localization in interacting quantum disordered systems attracts growing  interest because of its significance in quantum informatics \cite{PolkovnikovRevModPhys} where delocalization can reduce a quantum hardware performance, and in atomic physics \cite{PolkovnikovRevModPhys,HusePhen2,PreprintML1,PreprintML2} where many-body systems can be constructed and studied using cold neutral atoms in a magneto-optical trap.  

The single particle model is relevant at temperature approaching zero where the number of excitations in the whole system is small so their interaction can be approximately neglected. At finite temperature many-body interaction  makes the localization problem much more complicated for instance because it can initiate the irreversible energy transport stimulating the particle transport \cite{FleishmanAnderson1980}.  The single-particle localization breakdown due to  many-body interactions has been predicted in a variety of disordered quantum  systems including electrons in low dimensional  disordered metals ($d=1,2$) \cite{AronovReview,MirlinMBDel}, quantum defects in quantum crystals \cite{KMlocalization85} where the localization is expected in the absence of quenched disordering (see also \cite{Markus1,Markus2}), interacting two level systems in amorphous solids \cite{ab88Rv}  and molecular vibrations coupled by anharmonic interactions \cite{Leitner05}. It has been suggested in Refs.  \cite{Levitov0,Basko06} that the many body localization transition behaves similarly to the localization transition on the Bethe lattice because of similar phase space structures. Dramatic effect of the long-range many-body interaction destroying the localization transition in the infinite system even at arbitrarily strong disordering has been predicted in Refs. \cite{ab88Rv,abpreprint2d}.

The recent numerical investigations of many-body localization employing exact diagonalization methods \cite{HusePhen2,PreprintML2,OganesyanParamStudy,Reichmana,Reichmanb,IoffeFeigelman,HuseNewOrderParameter,HusePhen1,Pino} have provided  a high resolution view of the localization transition. Finite size scaling of these numerical results proves the existence of a many-body localization in  a one-dimensional system with a short-range interaction in the thermodynamic limit of an infinite system.  The specific nature of the localization transition and its relationship to the previously developed analytical theories remain unclear. For instance the level statistics used to identify the localization transition \cite{ShklShapiro}  behaves differently  in the many-body problem \cite{OganesyanParamStudy} compared to the Bethe lattice \cite{Bethe}. 
The investigation of the system with the long-range interactions \cite{PreprintML2} suggests the absence of localization transition for $1/R$ interaction leaving the situation with faster decreasing interactions inconclusive.

\begin{table}
 \caption{Analytical theory predictions for the critical interaction exponent $\alpha_{*}$, size $R_{*}$ and disordering $W_{*}$ corresponding to the localization transitions and the relaxation rate $k$ in the delocalized phase $N>N_{*}$ in the system with the long-range, $r^{-\alpha}$, spin-spin interactions, $\alpha < \alpha_{*}$. $\bar{V}=U_{0}n^{\frac{\alpha}{d}}$ stands for the interaction of neighboring spins.} 
 \label{tbl:scaling}
\centering
\begin{tabular}{|l|c|c||r|}
  \hline
   $\alpha_{*}$ & $N_{*}$ ($\alpha<\alpha_{*}$) & $W_{*}$ ($\alpha<\alpha_{*}$) & $k$ ($\alpha<\alpha_{*}$, $N>N_{*}$) \\
   \hline
   $2d$ & $\left(\frac{W}{\bar{V}}\right)^{\frac{d}{2d-\alpha}}$ & $\bar{V}N^{\frac{2d-\alpha}{d}}$ & $\frac{\bar{V}}{\hbar}\left(\frac{\bar{V}}{W}\right)^{\frac{\alpha+d}{2d-\alpha}}$ \\ 
   \hline  
  \end{tabular}
\end{table}

The aim of the present work is the numerical investigation of a many-body localization in systems with the long-range interaction decreasing with the distance $R$ as $R^{-\alpha}$  using the finite size scaling method. This method is very convenient to reveal the  expectations of analytical theory \cite{ab88Rv,PreprintML2} involving the lower constraint for the interaction law exponent $\alpha > \alpha_{*}=2d$ still permitting the localization in the infinite system and the power law size dependence of the critical disordering $W_{c} \propto N^{2d-\alpha}$ corresponding to the localization transition at $\alpha < \alpha_{*}$ where disordering $W$ is determined by the characteristic scale of a random potential acting on all spins. 
Moreover this and other scaling relationships (see Table \ref{tbl:scaling}) absent in the case of a short-range interaction can serve as a guideline for understanding of a very complicated many-body localization transition. 

The proposed study is also significant practically since the long-range interaction decreasing with the distance according to the power law inevitably exists between quasiparticles possessing charges or dipole, magnetic or elastic moments and it can significantly influence the localization transition similarly to a single particle case \cite{AndersonClassic,Levitov1,Levitov2} or even stronger \cite{ab88Rv}.

\begin{figure}[h!]
\centering
\includegraphics[width=3cm]{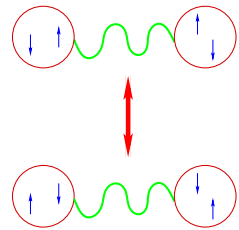}
\caption{ Two resonant pairs coupled by the many-body spin-spin interaction and the collective transition induced by that interaction.}
\label{fig:ResPairs}
\end{figure}

Below we briefly review the predictions of the analytical theory using the energy delocalization in an ensemble of interacting resonant pairs (Fig. \ref{fig:ResPairs}) as a precursor for a many-body delocalization for the arbitrarily system diemnsion $d$. Then the numerical results for $d=1$ are described and their consistency with the analytical theory and application to the anomalous relaxation of two level systems in amorphous solids are discussed.

We investigate the model of $N$ interacting spins $1/2$ placed  into equally spaced sites of a $d$-dimensional hypercube with the spatial density $n$. These spins are subjects to uncorrelated  random $z-$directional fields ($\phi_{i}S_{i}^{z}$) uniformly distributed within the domain $(-W/2, W/2)$. The interaction between spins $i$ and $j$ has two components $U_{ij}S_{i}^{z}S_{j}^{z}$ and $V_{ij}(S_{i}^{+}S_{j}^{-} + S_{i}^{-}S_{j}^{+})$ depending on  the distance as $u_{ij}/R_{ij}^{\beta}$, $u_{ij}/R_{ij}^{\alpha}$ with $\alpha=\beta$ and random sign interaction constants $u_{ij}, v_{ij} = \pm U_{0}$ as in the ``anisotropic'' interaction case of Ref. \cite{PreprintML2}.  The more general case $\alpha \neq \beta$ is interesting but more complicated  \cite{PreprintML2} and requires a special consideration. 

The strong coupling between spins takes place in resonant pairs where the change in the diagonal (Ising) energy  \cite{PreprintML2} $\mid\Delta_{ij}\mid=\mid \phi_{i}-\phi_{j}+\sum_{k\neq i,j}(U_{ik}-U_{jk})S^{z}_{k}\mid$ due to the flip-flop transition of spins $i$ and $j$ is smaller than the flip-flop interaction $V_{ij}$ (at strong disordering spin projections to the $z$ axis can be assumed well defined in the given quantum state). The probability of the formation of such pair by the interaction at some distance $R$ is given by $\frac{U_{0}}{WR^{\alpha}}$ (cf. Refs. \cite{ab88Rv,Levitov1}). Consequently 
the density of resonant pairs of a certain size $R$ (second spin can occupy the volume $R^d$) can be estimated as 
\begin{eqnarray}
n_{p}(R) \sim \frac{nR^{d}U_{0}}{WR^{\alpha}}=
n\frac{U_{0}n^{\frac{\alpha}{d}}}{W}
\left(\frac{1}{nR^{d}}\right)^{\frac{\alpha-d}{d}}. 
\label{eq:respairs}
\end{eqnarray}
Then the characteristic flip-flop interaction of resonant pairs (see Fig. \ref{fig:ResPairs}) can be expressed using their  interaction at the average distance between them, $n_{p}^{-1/d}$, as 
\begin{eqnarray}
V(R) \sim U_{0}n_{p}(R)^{\frac{\alpha}{d}}=U_{0}n^{\frac{\alpha}{d}}\left(\frac{U_{0}n^{\frac{\alpha}{d}}}{W}\right)^{\frac{\alpha}{d}}\left(\frac{1}{nR^{d}}\right)^{\frac{\alpha(\alpha-d)}{d^2}}. 
\label{eq:flflresp}
\end{eqnarray}
The energy delocalization within the subset of pairs of a certain size $R$ is expected if their coupling $V(R)$ exceeds 
the characteristic energy (disordering) of such pairs of the size $R$ given by 
\begin{eqnarray}
E(R) \sim \frac{U_{0}}{R^{\alpha}}=U_{0}n^{\frac{\alpha}{d}}\frac{1}{\left(nR^{d}\right)^{\frac{\alpha}{d}}}. 
\label{eq:PairEnergy}
\end{eqnarray}

Comparing behaviors of coupling strength Eq. (\ref{eq:flflresp}) and typical energy Eq. (\ref{eq:PairEnergy}) in the large size $R$ limit one can see  that at sufficiently small interaction exponent  $\frac{\alpha(\alpha-d)}{d^2}<\frac{\alpha}{d}$ the flip-flop interaction Eq. (\ref{eq:flflresp}) always exceeds the typical energy of pairs Eq. (\ref{eq:PairEnergy}). This inequality suggests that the  delocalization always takes place at sufficiently large system size if $\alpha<\alpha_{*}=2d$  \cite{ab88Rv,abpreprint2d}.

Below the threshold   ($\alpha < 2d$) one can expect that delocalization takes place for the resonant pairs of the size $R\sim R_{*}$ where their flip-flop interaction Eq. (\ref{eq:flflresp}) approaches their energy Eq. (\ref{eq:PairEnergy}) ($V(R_{*})=E(R_{*})$). This size given in the table \ref{tbl:scaling} estimates a minimum system size where the delocalization takes place.  Consequently the minimum number of interacting spins needed for delocalization can be expressed using the density of spins $n$ and critical radius $R_{*}$ as $N_{*}\sim nR_{*}^{d}$ (see Table \ref{tbl:scaling}). 

Using the spectral diffusion consideration in Refs.  \cite{ab93,ab88Rv,ab88}  one can also estimate the spin relaxation rate $k_{1}$ in the delocalization regime $N>N_{*}$ as the ratio of a typical energy transfer rate between resonant pairs  $k_{*}\approx \frac{U_{0}}{\hbar R_{*}^{\alpha}}$ and a relative fraction of resonant pairs $x_{*}=\frac{n_{p}(R_{*})}{n}$ (see final expression in Table \ref{tbl:scaling}). 

The dynamics in the delocalization domain can be characterized by the spin relaxation rate which is the average time between the spin irreversibly switches between its $\pm 1/2$ states. We expect that the transport of spins can be described as a spin diffusion with the coefficient given by $D \sim kR_{*}^2$  although this expectation needs further numerical verifications considering possible diffusion coefficient reduction (subdiffusion) with increasing the system size \cite{Reichmana} or because of the opposite effect of Levy flights due to the long-range interaction   \cite{Altshuler2dDipole} (superdiffusion).

In the numerical analysis  we set $U_{0}=\hbar=n=1$ expressing disordering in $U_{0}$ units, time in $\hbar U_{0}^{-1}$ units and relaxation rates in $U_{0}/\hbar$ units. Interspin distance $R_{ij}$ is taken as a minimum distance alone the closed chain ($R_{ij}=\min(|i-j|, |N-|i-j|)$). 
To characterize the many-body localization we use the ergodicity parameter introduced in Ref.  \cite{BGold} as the spin-spin local correlation function $Q(t)=4<S^{z}(t)S^{z}(0)>'$ taken in the infinite time limit 
\begin{eqnarray}
Q=4\left<\mid\left<\beta|S_{i}^{z}|\beta\right>\mid^{2}\right>'. 
\label{eq:QExpInf}
\end{eqnarray}
The symbol $<...>'$ means that the eigenstates $|\beta >$ are collected from the narrow energy domain $(E-\Delta, E+\Delta)$ where $E=0$ is the energy of interest and the energy range $\Delta$  has been chosen as  $\Delta E=0.04W\sqrt{N}$ to have the change in the many-body density of states less than by $1$\%. This consideration corresponds to the infinite temperature limit often used to study the many-body localization \cite{PreprintML2,OganesyanParamStudy,Reichmana}.

The ergodicity parameter in Eq. (\ref{eq:QExpInf}) is expected to be finite in the localization regime and to approach zero in the delocalized state of an infinite system because of the long time  correlation decoupling $Q(t)=<S^{z}(t)S^{z}(0)>'\rightarrow <S^{z}(t)><S^{z}(0)>$ and zero average spin projection at infinite temperature.  The time dependent local correlation function $Q(t)$ characterizes the spin relaxation. 

Calculations of ergodicity parameter and time dependent spin-spin correlation function were performed using Matlab software on the linux cluster available through Tulane University Center for Computational Science. Random Hamiltonians have been generated for each specific exponents $\alpha=1, 1.5, 1.75, 2, 2.25, 2.5, 3, 10$, disordering $2< W <150$, and total numbers of spins $N=8, 10, 12, 14$. The conserving projection of  the total spin to the $z$-axis has been always set to $0$ in agreement with the assumption of an infinite temperature. Then the ergodicity parameter has been calculated diagonalizing each Hamiltonian. The results have been averaged over a  sufficiently large number of realizations chosen to make the relative error of the estimate less than $1\%$. The more complicated calculations of time dependent correlation function has been performed only for the practically significant case $\alpha=d=1$ and short range limit $\alpha=10$. The results and their analysis are presented below. 

\begin{figure}[h!]
\subfloat[]{\includegraphics[scale=0.15]{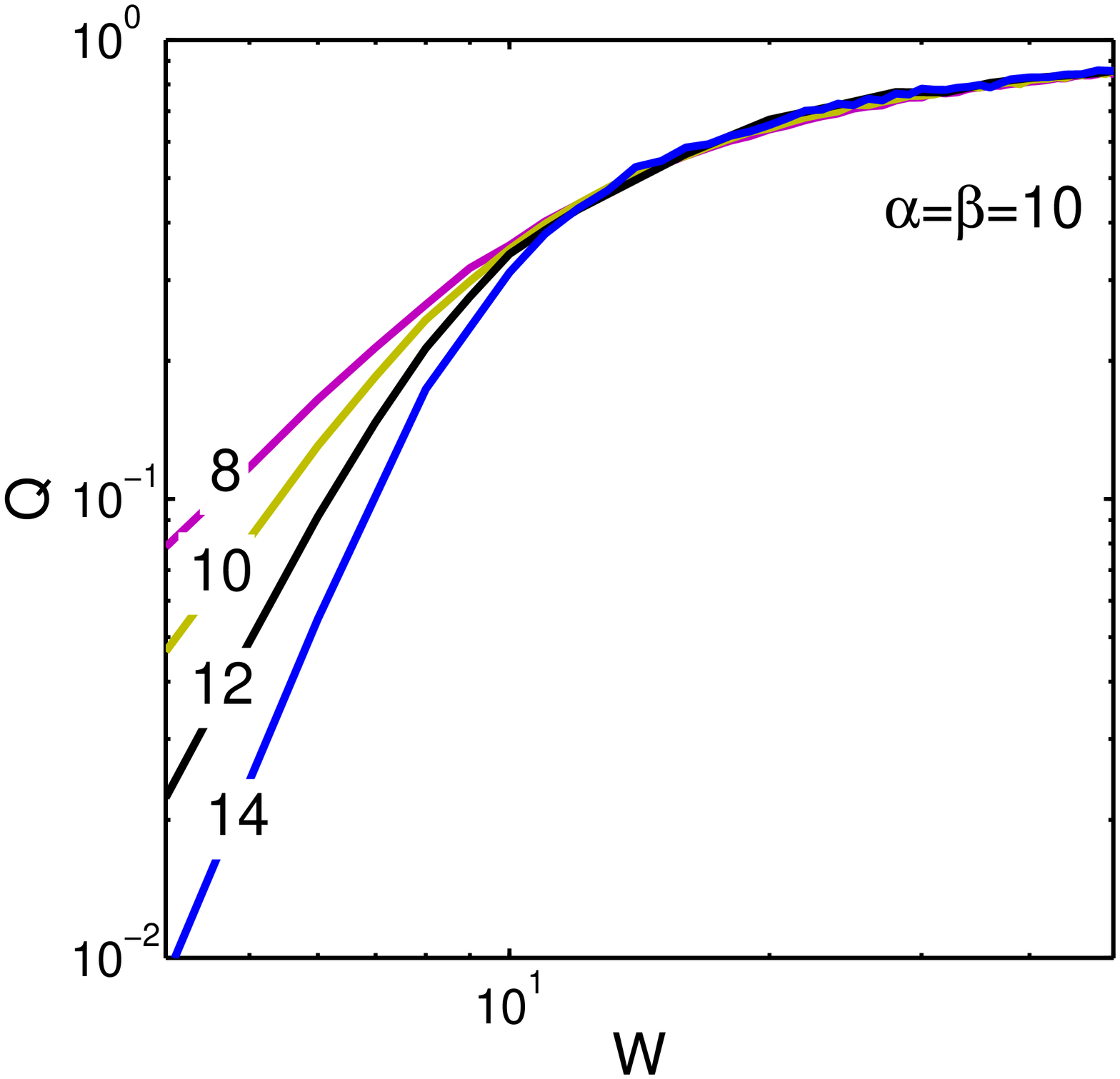}}
\subfloat[]{\includegraphics[scale=0.15]{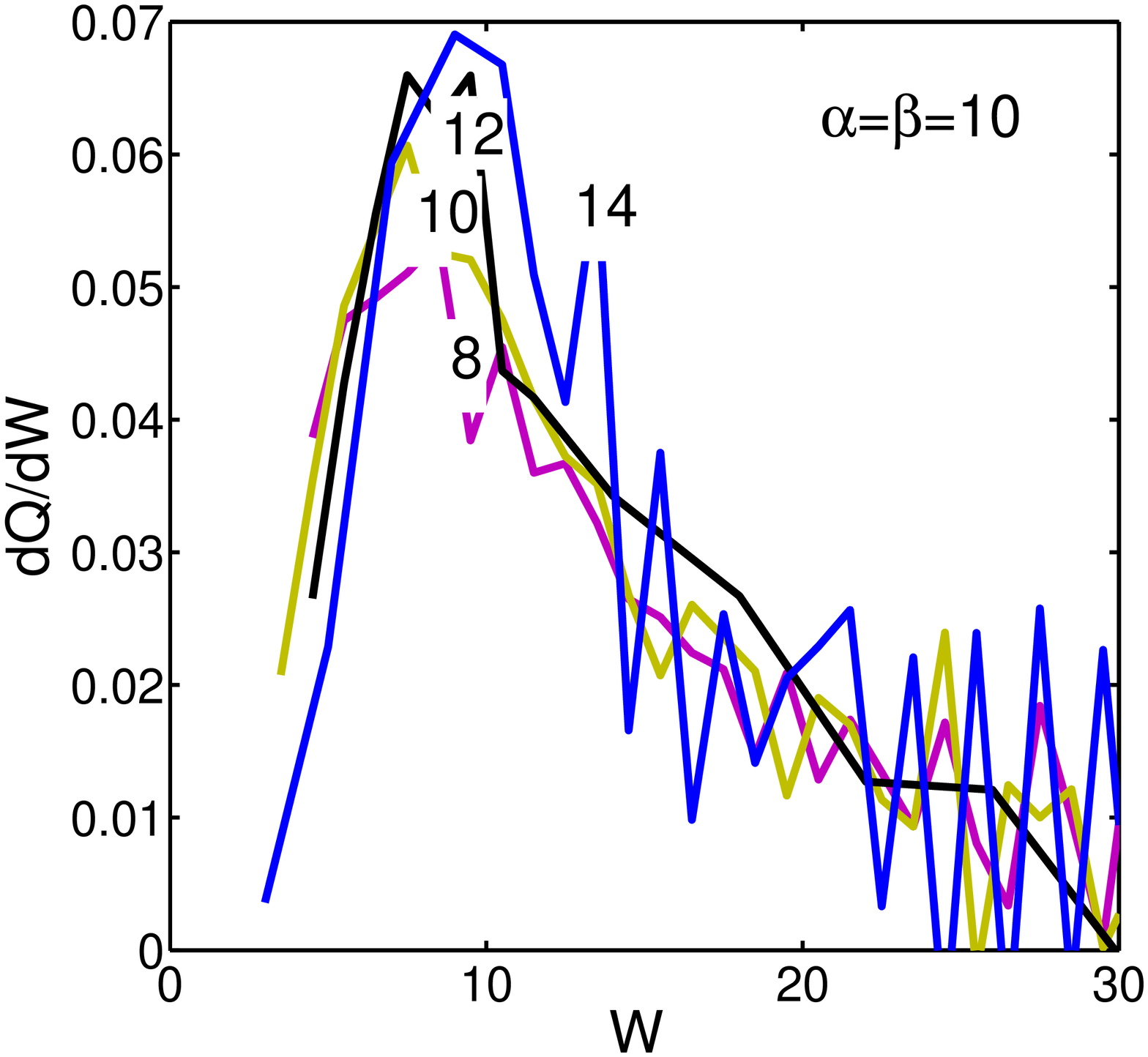}}\\
\subfloat[]{\includegraphics[scale=0.15]{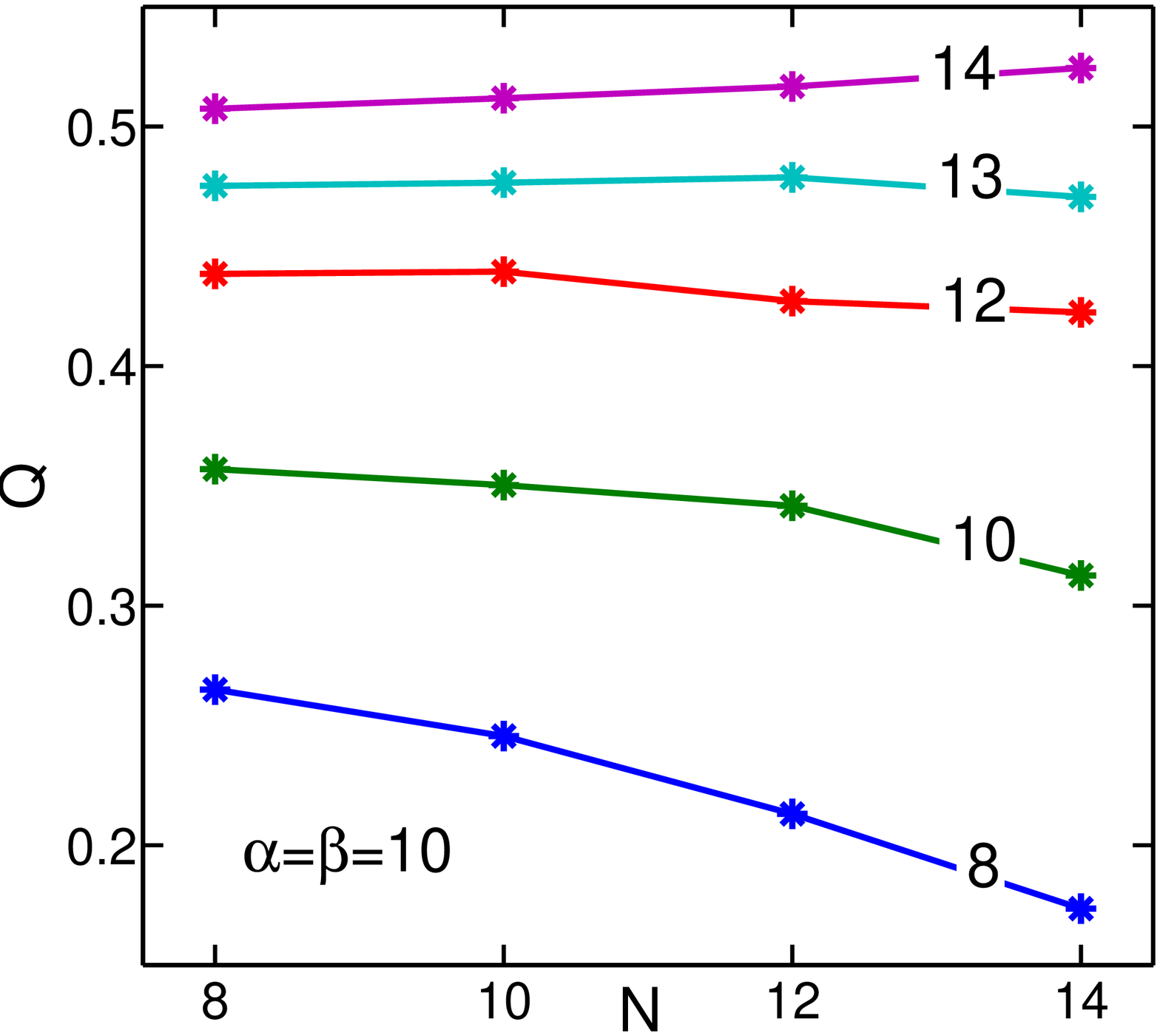}}
\subfloat[]{\includegraphics[scale=0.15]{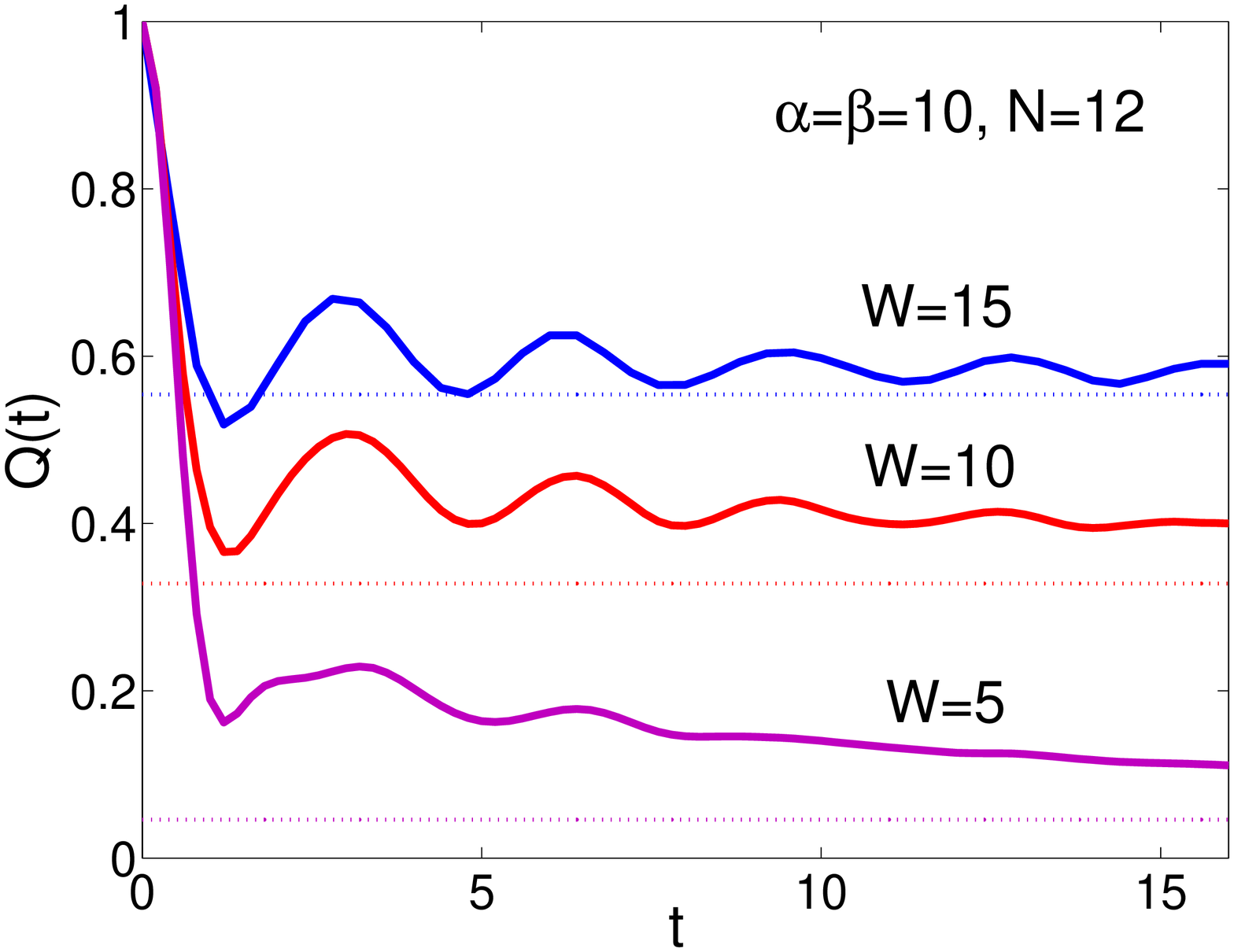}}
\caption{\small Localization and dynamics of spin system with a short-range interaction $\alpha=\beta=10$ ({\bf a, b}) Ergodicity parameter $Q$ (a) and its derivative $dQ/dW$ (b) vs. disordering; the chain length is shown at each line.  ({\bf c}) Finite size scaling for ergodicity parameter with disordering shown at each line.  ({\bf d}) Time dependence of spin-spin correlation function $Q(t)$ for a system of $12$ spins; disordering is shown at each line. Dashed lines indicate the limit of an infinite time.}
\label{fig:a10}
\end{figure} 

We begin with the consideration of the short-range limit of $\alpha=\beta=10$ (see Fig. \ref{fig:a10}) where the size-independent many-body localization takes place  \cite{HusePhen2,PreprintML2,Pino}.  Indeed  at $W>10$ the ergodicity parameter is almost size independent (Fig. \ref{fig:a10}a).  The maximum of its derivative with respect to disordering (inflection point, see Fig. \ref{fig:a10}b) estimated using finite differences seems to be not sensitive to the size as well, though the accuracy of the derivative estimate is too low to determine the threshold. The consideration of the ergodicity parameter size scaling at constant disordering (Fig. \ref{fig:a10} c) similar to that in Refs.  \cite{HusePhen2,PreprintML2} gives the estimate of the critical disordering $W_{c} \approx 12 \pm 2$ in accord with the earlier work  \cite{PreprintML2}. We will use the results for the ergodicity parameter to characterize  many-body localization transitions in the case of the long-range interactions employing data rescaling as described below.  The time-dependent behavior (Fig. \ref{fig:a10}d) will be discussed later in comparison to that for the long-range interaction $\alpha=\beta=1$. 

\begin{figure}[h!]
\centering
\subfloat[]{\includegraphics[scale=0.15]{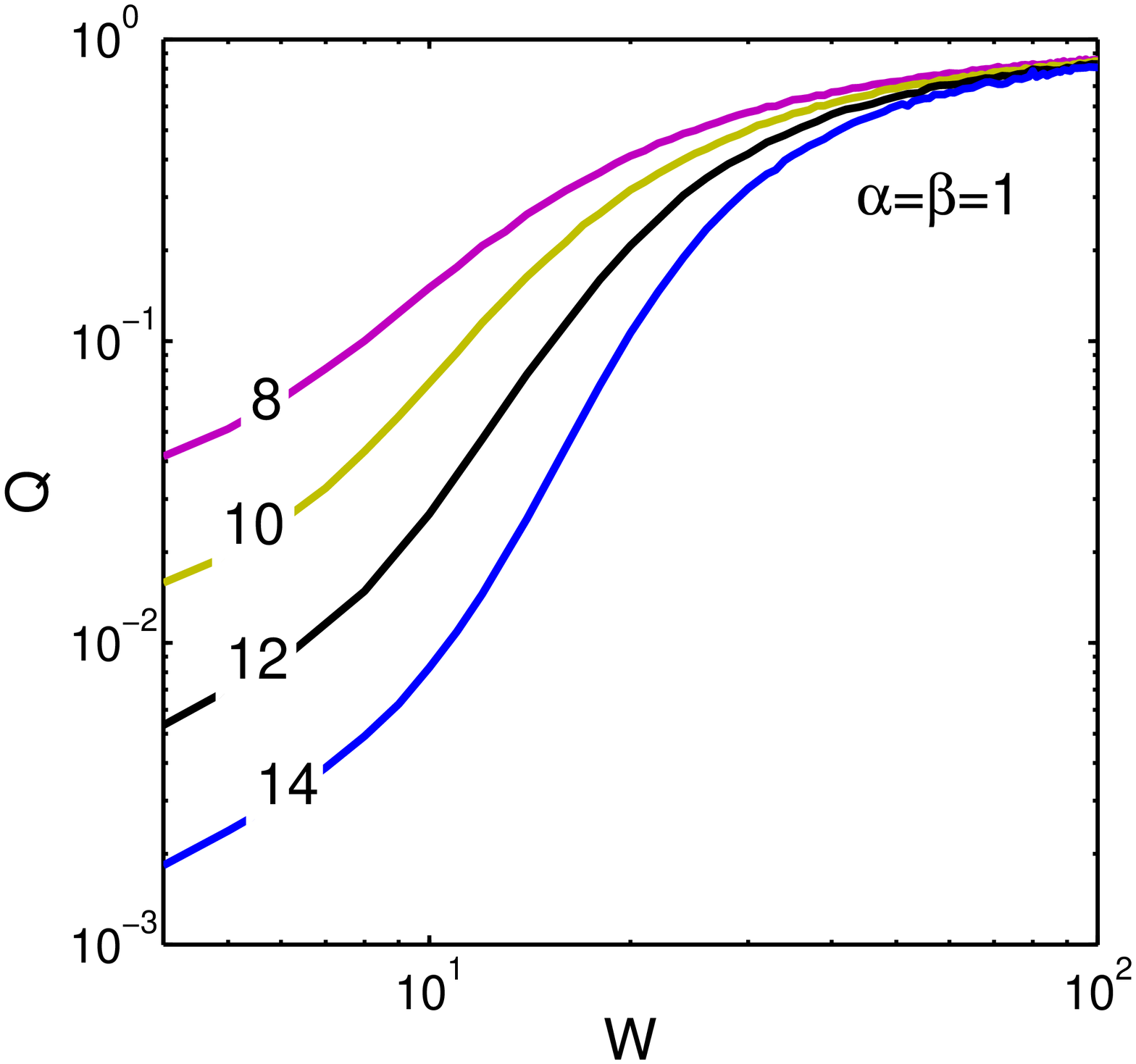}}
\subfloat[]{\includegraphics[scale=0.15]{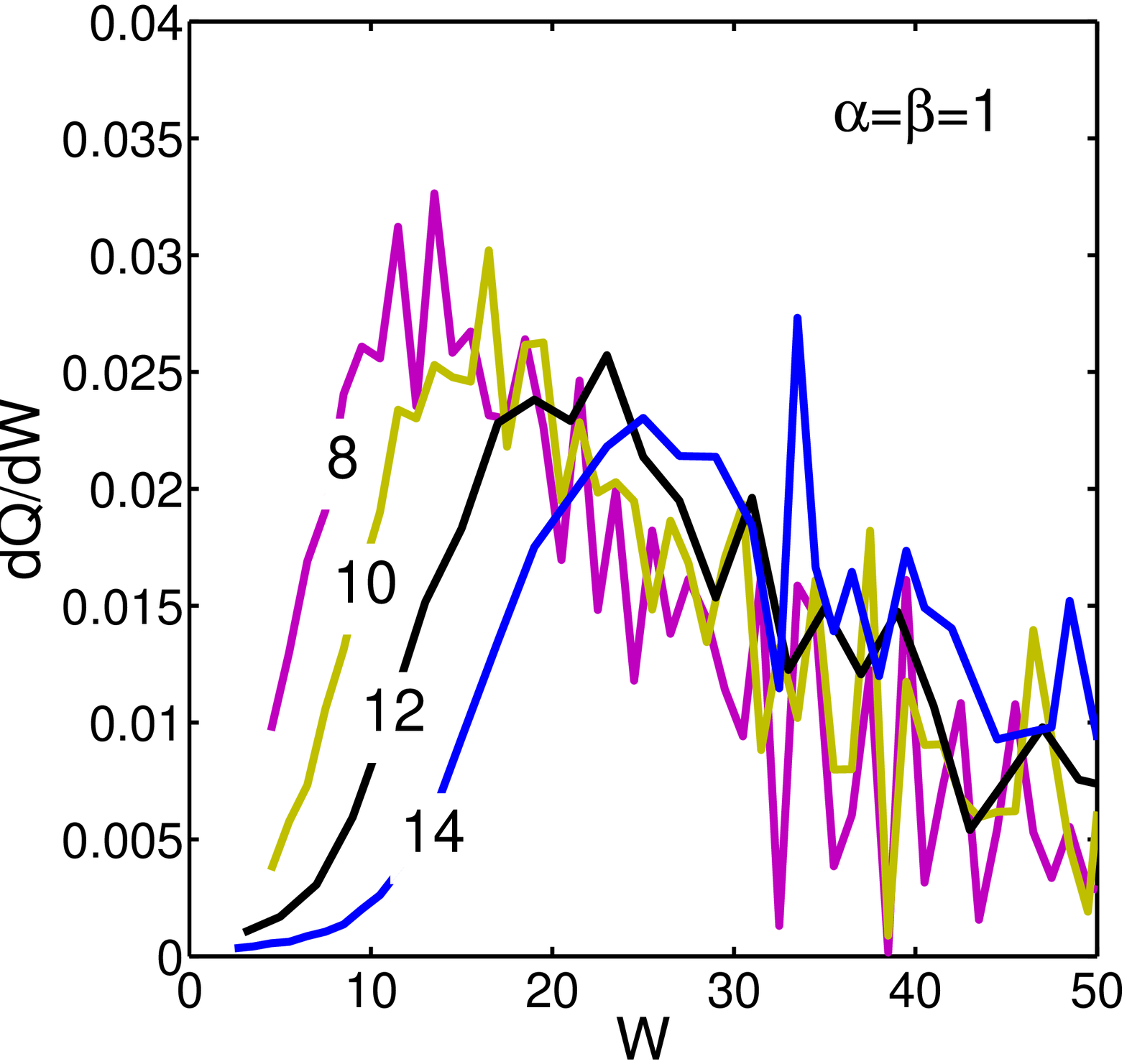}}\\
\subfloat[]{\includegraphics[scale=0.15]{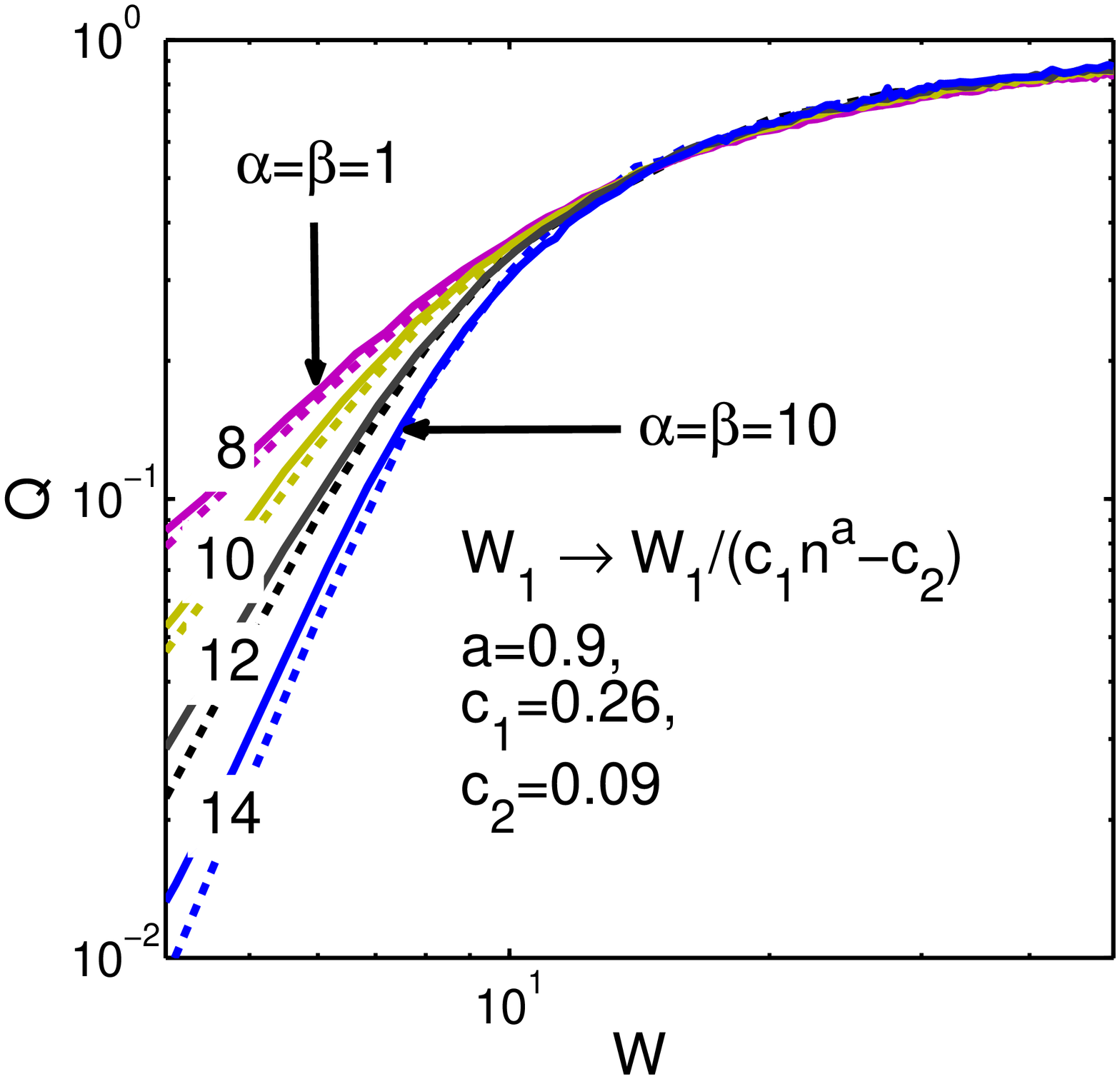}} 
\subfloat[]{\includegraphics[scale=0.145]{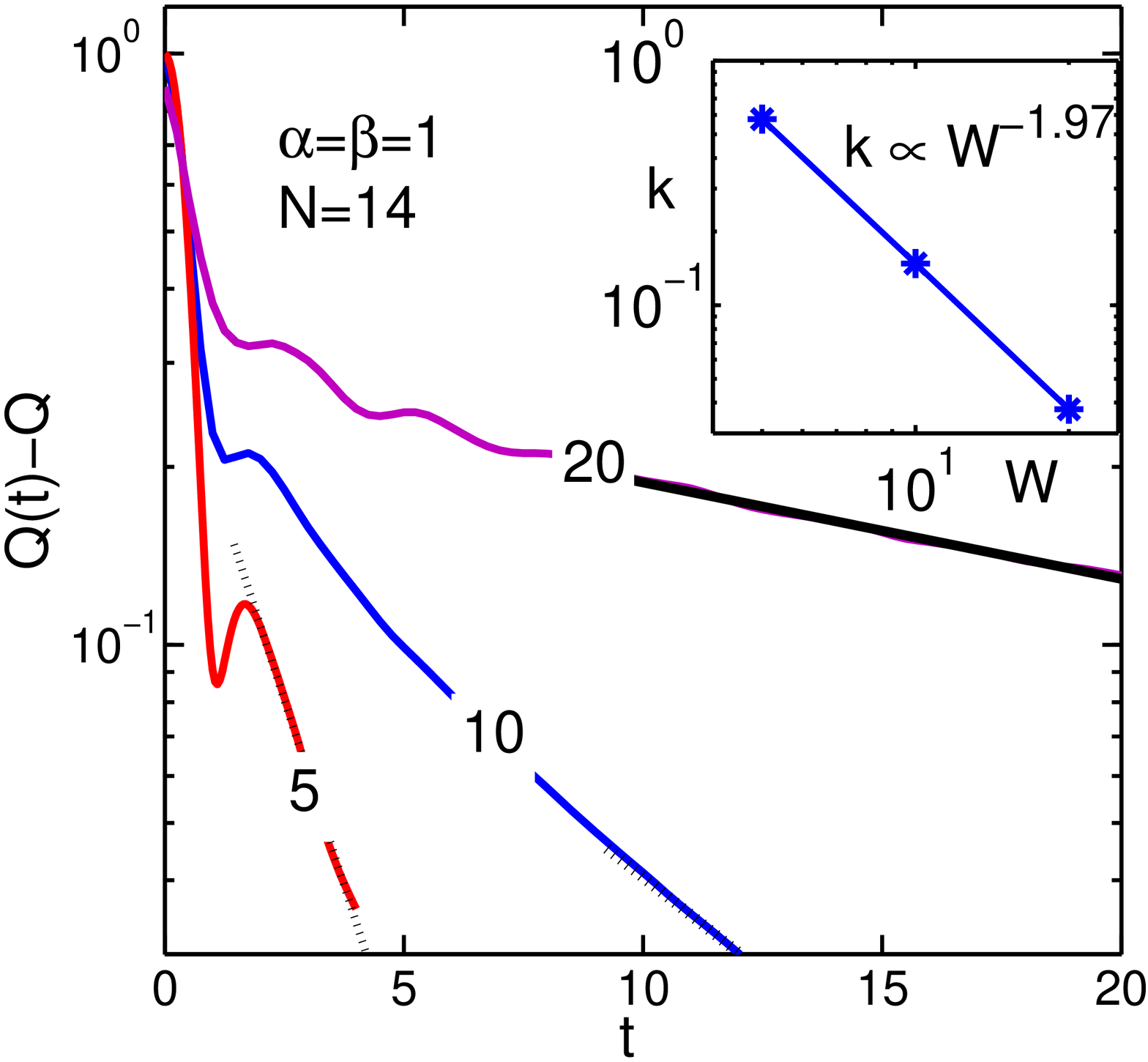}}
{\caption{\small Results for the long-range interaction $\alpha=\beta=1$ ({\bf a, b}) Ergodicity parameter $Q$ (a) and its derivative $dQ/dW$ (b) vs. disordering; the chain length is shown at each line. ({\bf c}) Ergodicity parameter $Q$ vs. rescaled disordering  compared with the short-range results shown by dashed lines.  ({\bf d}) Time dependence of the spin-spin correlation function $Q(t)$ for the system of $N=14$ spins; disordering is shown at each line; solid straight lines indicate the exponential data fit $e^{-kt}$. Inset shows the dependence of the relaxation rate $k$ on disordering.}
\label{fig:a1}}
\end{figure}  

The results of calculations for the long-range interaction with the smallest exponents $\alpha=\beta=1$ are shown in Fig. \ref{fig:a1}. This regime ($\alpha=\beta=d$) is of a great interest because in $3D$ it describes dipolar, elastic or magnetic dipolar interactions.  
Since  the expected qualitative behaviors are sensitive only to the ratio $\alpha/d$ (see Table \ref{tbl:scaling}) the results for $\alpha=\beta=1$ for $d=1$ can be used for the case $\alpha=\beta=3$ in $3D$. 

The changes in behaviors of ergodicity parameter and its derivative with the system size clearly indicate the shift of a many-body localization transition  towards larger random potentials with growing the system size in a full accord with the analytical theory expectations (Table \ref{tbl:scaling}) and the recent work  \cite{PreprintML2}. Consider the dependence of the localization transition on the system size. This is a difficult task because the transition point is not defined clearly and the range of the considered system sizes is not very broad to rely on inaccurate data for the inflection point. Noticing similarities in shapes of the graphs for ergodicity parameter in a short-range (Fig. \ref{fig:a10}.a) and long-range (Fig. \ref{fig:a1}.a) regimes ranging from large  disordering down to the estimated transition point and even below it  we rescale random potentials depending on the size for the long-range interaction to attain the best match between long-range and short-range regimes for each size. The rescaling was probed using the power law size dependence 
\begin{eqnarray}
W \rightarrow \frac{W}{(c_{1}n^{a}-c_{2})\cdot {\rm sign}(a)}. 
\label{eq:rescaling}
\end{eqnarray}
(the additional constant term $c_{2}$ has been introduced to account for possible finite size effects. The optimum set of fitting parameters $a, c_{1}, c_{2}$ has been obtained  using the Monte Carlo procedure (see  \cite{abEcho}) developed earlier for the fitting of multiple experimental data sets. 

The results of fitting and the optimum fitting parameters are shown in Fig. \ref{fig:a1}.c. The graphs for the rescaled long-range and short-range interactions look almost identical from well below the estimated transition point at $W_{c} \approx 12$ up to strong disordering corresponding to the localization.    It is remarkable that the scaling exponent $a=0.9$ is very close to the analytical theory prediction $a=2d-\alpha=1$ (see Table \ref{tbl:scaling}). If we neglect this minor difference then the critical disordering needed for the many-body localization for $\alpha=\beta=1$ behaves with the system size as $W_{c,1}=3.1 \cdot N$. 

\begin{figure}[h!]
\centering
\subfloat[]{\includegraphics[scale=0.15]{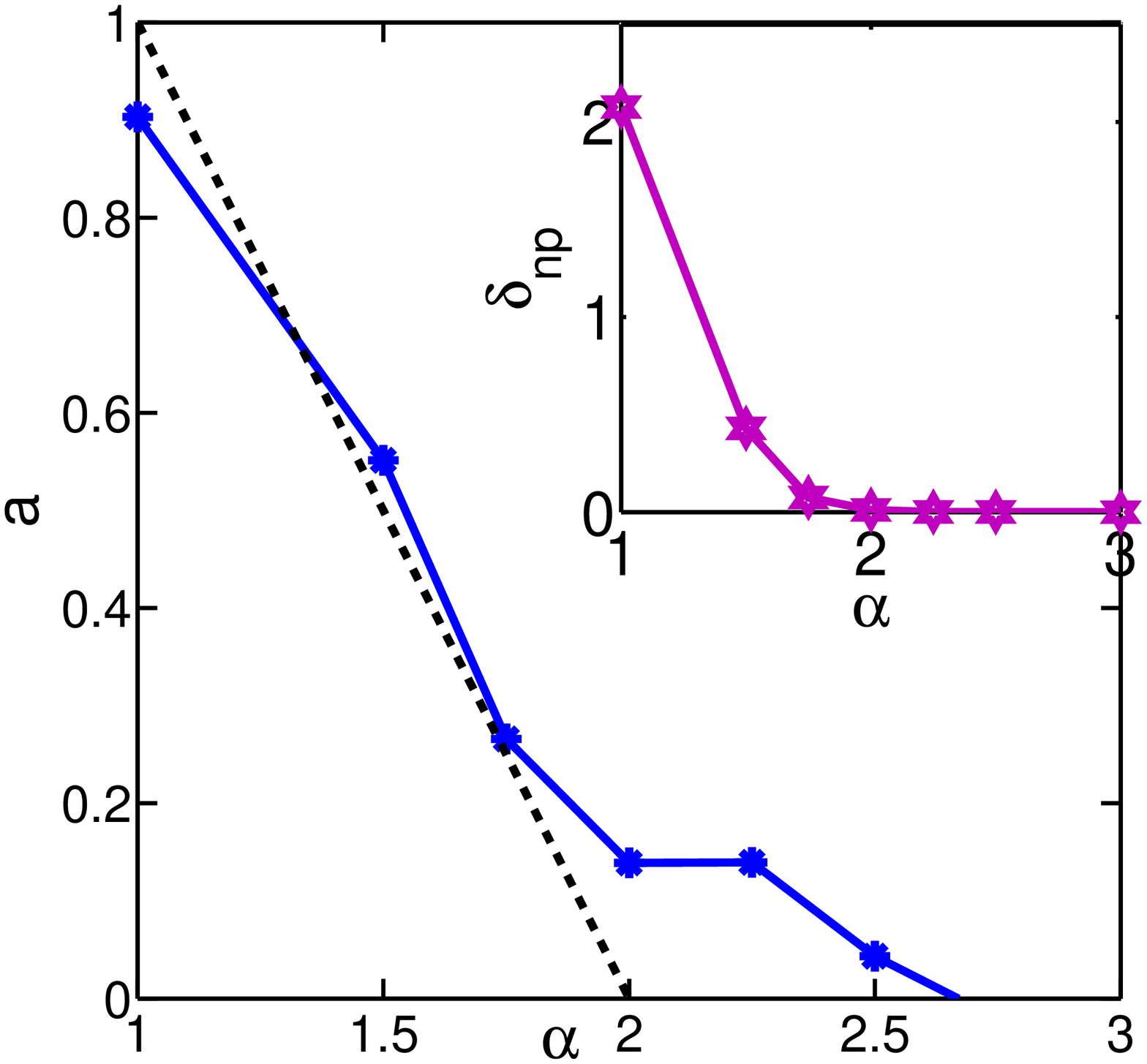}}
\subfloat[]{\includegraphics[scale=0.15]{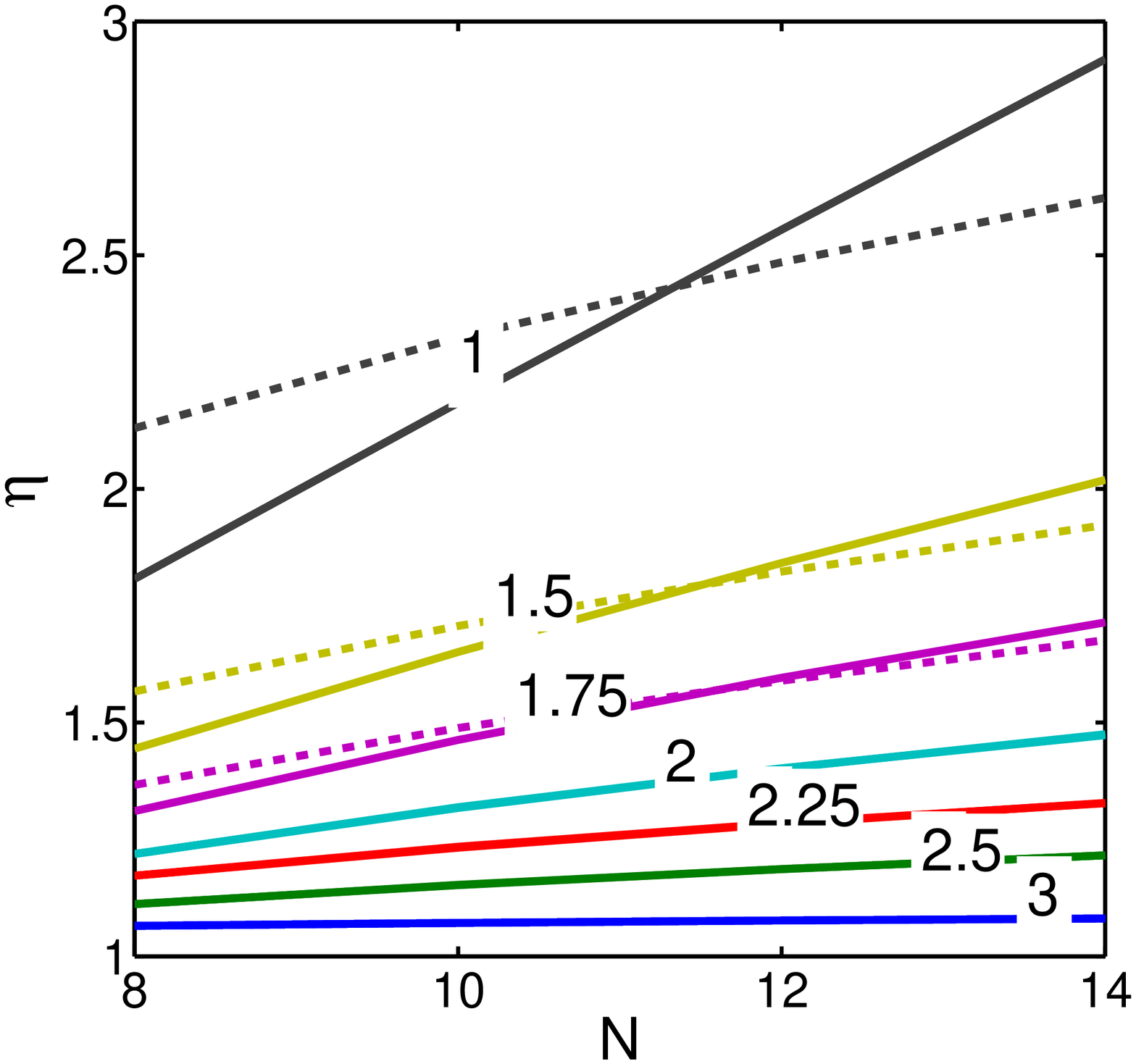}}
{\caption{\small Data rescaling vs. interaction exponent $\alpha$ ({\bf a}) Scaling exponent $a$ for the optimum fit using Eq. (\ref{eq:rescaling}). Straight dashed line shows theory predictions. Inset shows the relative difference between the standard deviations for the given fit and the fit using a negative exponent $a$. ({\bf b}) Rescaling of disordering obtained using  optimum fits with positive (straight lines) and negative (dashed lines) exponents $a$. Interaction exponent $\alpha$ is shown at each line.}
\label{fig:ExpErr}}
\end{figure}

The same fitting procedure has been applied to other considered interaction exponents. The results for the optimum rescaling are given in Fig. \ref{fig:ExpErr}. For interaction power law exponents $\alpha=1, 1.5, 1.75$ where delocalization is expected the scaling exponents $a=0.9, 0.55, 0.26$  have been found for the critical disordering (see Eq. (\ref{eq:rescaling})). They  are consistent with the analytical theory predicting $a=2d-\alpha=1, 0.5, 0.25$ (see Table \ref{tbl:scaling}). We attempted to rescale the ergodicity parameter assuming negative exponents $a$ corresponding to the saturation of the size dependence and, consequently, localization in the infinite size limit. These attempts for $\alpha < 2$ result in much worse fits as shown in the inset in Fig. \ref{fig:ExpErr}.a.

In the threshold case $\alpha=\beta=2d=2$ the disordering scaling is obtained in the form $2.57N^{0.14}-2.26$. In the size range of interest this dependence cannot  be distinguished from  the logarithmic dependence  consistent with the natural expectations for the threshold regime where the number of resonant interactions grows logarithmically with the system size  \cite{Levitov1}. 

The comparable or smaller scaling exponents ($a=0.1395, 0.0432, -0.0844$) are obtained for faster decreasing  interaction ($\alpha=2.25, 2.5, 3$, respectively). The relative weight of a size-dependent  term decreases with increasing $\alpha$ as seen in Fig. \ref{fig:ExpErr} b. Yet the fit in the present size range suggests the weak, nearly logarithmic size dependence of rescaling factor  even at $\alpha>2d$. This dependence can be saturated at larger sizes or can be due to another delocalization mechanism weaker than the one considered within the analytical theory. This problem can be clarified extending the analysis to   larger sizes $N$. 


Finally we discuss the relaxation for $\alpha=\beta=1$ shown in Fig. \ref{fig:a1}.d and compare it to the short-range interaction case (Fig. \ref{fig:a10}.d).  Two stages of dynamic behaviors  can be distinguished in both cases similarly to the $\alpha$ and $\beta$ relaxations in the mode coupling theory of a glass transition \cite{Goetze,Reichmanc}. First stage includes the fast reduction of $Q(t)$ during the time $t_{\beta}\approx 1$. In the second stage the correlation function slowly approaches its  ``equilibrium'' $Q=Q(\infty)$. This stage is accompanied by some oscillations  expressed stronger in the short-range interaction case, which makes it difficult for a quantitative analysis.  We suggest that the relaxation in the first stage is associated with virtual flip-flop transitions in infrequent resonant pairs occurring mostly between adjacent spins. Accordingly, their characteristic time is determined by the inverse coupling strength $\hbar/U_{0} \approx 1$. In our opinion the delocalization occurs in the second stage. In the case of the long-range interaction this relaxation should be associated with the collective many-body dynamics of resonant pairs (Fig. \ref{eq:respairs}). In this regime the spin fluctuation correlation function $Q(t)-Q(\infty)$ can be fitted well by a single exponent (see Fig. \ref{fig:a1}.d). The extracted relaxation rate shown in the inset depends on disordering as $k \propto W^{-1.97}$ in the excellent agreement with the dependence $W^{-2}$ predicted by the analytical theory (see Table \ref{tbl:scaling}). The full relaxation rate can be expressed as 
\begin{eqnarray}
k\approx 15\frac{(U_{0}n)^3}{\hbar W^2}. 
\label{eq:relax}
\end{eqnarray}
 
A similar qualitative behavior of a relaxation rate is expected in any $d-$dimensional system with the interaction decreasing with the distance as $R^{-d}$ since the analytical theory predictions are sensitive to the ration $\alpha/d$ only. The numerical prefactor can be very different though because the interaction $U_{0}n$ comes from counting the number of resonances and it might have the unit sphere area factor $a_{d}$ in its definition. This factor  is equal to $2$, $2\pi$ and $4\pi$ for $d=1, 2$, and  $3$, respectively. A multiplication of Eq. (\ref{eq:relax}) by $a_{d}^3$ results in the numerical prefactor of order of $10^3$ in $3D$. This expectation can clarify the long standing problem of the nature of the anomalous relaxation of two level systems (TLS's) in dielectric glasses  \cite{ab88Rv,abEcho} showing linear temperature dependence at low temperature instead of the $T^3$ dependence due to the TLS-phonon interaction. 

Indeed our model can describe interacting TLS's setting disordering to the thermal energy, $W \rightarrow k_{B}T$, as the typical energy of a ``mobile'' thermal TLS having tunneling amplitude and energy of order of $k_{B}T$.  The  interaction constant $U_{0}$ should be replaced  with the characteristic  interaction energy of nearest ``thermal'' TLS's, $U_{0}n \rightarrow k_{B}TP_{0}U$, where $P_{0}$ is TLS density of states  \cite{ab88Rv}) and $U$ is their $1/R^3$ interaction constant. Then the analytical theory (Table \ref{tbl:scaling}) predicts the linear temperature dependence similarly to Ref.  \cite{ab88Rv} while the prefactor is three order of magnitude smaller than the one observed experimentally (see Ref.  \cite{abEcho} and references therein). The suggested arguments about a prefactor of order of $1000$ can resolve this discrepancy. 


Thus the finite size scaling for the many-body localization problem with the long-range interaction is consistent with the analytical theory and the results can help to interpret the anomalous the low temperature relaxation in amorphous solids. 


This study has been stimulated by the recent work \cite{PreprintML2} investigating a many-body localization  with the long-range interaction. The author acknowledges Louisiana EPSCORE LA Sigma and LINK Programs for the support and Kevin Osborn for useful suggestions.

\end{document}